\documentstyle[epsfig]{elsart}

\newcommand{\be}{\begin{equation}}
\newcommand{\ee}{\end{equation}}
\newcommand{\beq}{\begin{eqnarray}}
\newcommand{\eeq}{\end{eqnarray}}

\begin{document}

\begin{frontmatter}
\title{Reactions $\gamma\gamma\to\pi\pi$ and $\gamma\gamma\to K \bar
K$:\\ the $(IJ^{PC}=00^{++})$-wave spectra at
$E_{\gamma\gamma}^{(c.m.)} \sim 300- 1900$ MeV}
\author{A.V.Anisovich and V.V.Anisovich}
\address{St.Petersburg Nuclear Physics
Institute, Gatchina, 188350, Russia}

\begin{abstract}
We calculate the $00^{++}$-wave two-meson spectra for the reactions
$\gamma\gamma\to\pi^0\pi^0$,
$\gamma\gamma\to\pi^+\pi^-$,
$\gamma\gamma\to K^0\bar K^0$ and
$\gamma\gamma\to K^+ K^-$
on the basis of:
(i) the results of the $K$-matrix analysis of the $00^{++}$
amplitudes for the reactions $\pi\pi \to \pi\pi$, $K\bar K$,
$\eta\eta$, $\eta\eta'$, $\pi\pi\pi\pi $, and
(ii) recently developed method for calculation of
the decay amplitudes
$f_0(1^3P_0q\bar q)\to \gamma\gamma$,
$f_0(2^3P_0q\bar q)\to \gamma\gamma$.
The reconstructed $\pi\pi $ and $K\bar K$ spectra
can be used as a guide for the extraction of
 partial widths of scalar/isoscalar resonances
$f_0(980)\to \gamma\gamma$,
$f_0(1300)\to \gamma\gamma$,
$f_0(1500)\to \gamma\gamma$,
$f_0(1750)\to \gamma\gamma$. As follows from our calculations,
the resonances $f_0(980)$ and $f_0(1500)$ in the $\pi^+\pi^-$ and
$\pi^0\pi^0$ spectra reveal themselves as dips.
\end{abstract}
\end{frontmatter}

\section{Introduction}

The reactions $\gamma\gamma\to mesons$
provide valuable information on the quark/gluon structure of
mesonic states, see \cite{barnes} and references
therein.  These reactions were intensively studied during the last
decade (see, for example, Refs. \cite{a0f0,f0,a0,PDG}), and their
investigations are carried on.

In the present paper we discuss the $00^{++}$-spectra in the two-meson
production processes
$\gamma\gamma\to\pi^0\pi^0$, $\gamma\gamma\to\pi^+\pi^-$,
$\gamma\gamma\to K^0\bar K^0$ and
$\gamma\gamma\to K^+ K^-$ where the production of
the scalar/isoscalar resonances provides a possibility to measure
partial widths  $f_0\to \gamma\gamma$.

Recently, the $\gamma\gamma$-widths were calculated for
scalar mesons, members of the $1^3P_0 q\bar q$
and $2^3P_0 q\bar q$ nonets \cite{aabn}. In the $10^+$-channels,
where scalar/isovector resonances, $a_0$, are seen, the background
is comparatively small.
However, for scalar/isoscalar resonances
there is a strong
interference of the resonance signal with a background,
that is quite analogous to the situation observed in the
analysis of the $\pi\pi \to \pi\pi$
$00^{++}$-wave at $M_{\pi\pi} \sim 600$-$1900$ MeV
\cite{aps,hep}. Because of that, when discussing the scalar/isoscalar
sector and $\gamma\gamma$ partial widths of the  $f_0$-states, we
should reconstruct both resonance and
background contributions.

There are two types of processes
which determine the prompt production of
mesons in the reactions $\gamma\gamma\to\pi\pi$ and $\gamma\gamma\to K
\bar K$ (see Fig. 1 where the diagrams for the reaction
$\gamma\gamma\to\pi^+\pi^-$ are shown): \\
(i) direct $s$-channel resonance production, Fig. 1a, and \\
(ii) non-resonance production (the main contribution, one can guess,
is due
to diagrams with the $t$- and $u$-channel
  pole singularities  located near the physical region, Fig. 1b). \\
In addition, a large background in the $00^{++}$-channel ultimately
brings out a significant final state interaction effect.

The $K$-matrix approach provides an appropriate technique for taking
into account both the prompt production processes  and those
 followed by the final state meson interaction. In Refs.
\cite{aps,hep}, the $K$-matrix
amplitude was restored for the processes
$\pi\pi \to \pi\pi$, $K\bar K$,
$\eta\eta$, $\eta\eta'$ and  $\pi\pi\pi\pi $
in the mass region $M_{\pi\pi} \leq 1900$ MeV: just this amplitude
is necessary for final state interactions
in the processes $\gamma\gamma  \to two \; mesons$.

In the next Section we write down the
$K$-matrix amplitudes for the processes
$\gamma\gamma\to\pi\pi$ and $\gamma\gamma\to K \bar K$.
In Section 3 the $K$-matrix elements for the prompt processes shown in  
Figs. 1a and 1b are presented. We discuss the results in Section 4, 
and a brief conclusion is given in Section 5.

\section{$K$-matrix amplitudes for
$\gamma\gamma\to\pi\pi$ and $\gamma\gamma\to K \bar K$}

Here the formulae are given for the process
$\gamma\gamma\to\pi^+\pi^-$; the amplitudes for other processes under
consideration are written in a similar way.

The amplitude  for the production of $S$-wave pions, with
$I=0$, in the reaction $\gamma\gamma\to\pi^+\pi^-$ reads:
\be
A_{\alpha ,\beta}(\gamma\gamma\to\pi^+\pi^-)= e^2
g^{\perp\perp}_{\alpha\beta}\widetilde K_{\gamma\gamma, a}
\left (\frac{{\bf I}}
{ {\bf I}-i\hat{\rho}\hat K}\right )_{a,\pi^+\pi^-}.
\label{1}
\ee
The indices  ($\alpha,  \beta$) refer to the photon polarization,
and the metric tensor
$g^{\perp\perp}_{\alpha\beta}$
works in the space perpendicular to the momenta of colliding photons.
$\widetilde K_{\gamma\gamma, a}$ is the
$K$-matrix element for the prompt transition
$\gamma\gamma \to a$, and the index $a$ runs over seven states:
$\pi^0\pi^0$, $\pi^+\pi^-$, $K^0\bar K^0$,$K^+ K^-$,
$\eta\eta$, $\eta\eta'$ and  $\pi\pi\pi\pi $.
The factor
$({\bf I}-i\hat{\rho}\hat K)^{-1}$
is responsible for the final state
interaction of mesons: ${\bf I}$ is the unity matrix,
$\hat\rho$ is the diagonal phase space
matrix, $\hat\rho=diag(\rho_1,\rho_2,\rho_3,...)$,
where $\rho_a$ is the phase space of the state $a$, and the
matrix elements $K_{ab}$ refer to meson-meson
interaction in the channels
$\pi^0\pi^0$, $\pi^+\pi^-$, $K^0\bar K^0$,$K^+ K^-$,
$\eta\eta$, $\eta\eta'$ and  $\pi\pi\pi\pi $.

\subsection{$K$-matrix representation of the $00^{++}$ wave
meson-meson amplitude}

In the c.m. energy region $\sqrt s \leq 1900$ MeV,
the $K$-matrix amplitude for  meson-meson
interaction in the channels
$\pi\pi $, $K\bar K$,
$\eta\eta$, $\eta\eta'$ and  $\pi\pi\pi\pi $
has been found in \cite{aps}  on the basis of a sample of
data from Refs. \cite{gams,bnl,cbc}.
The matrix elements were parametrized
as a sum of the pole and background  terms:
\be
K_{ab}(s)=
 \sum_\alpha \frac{g^{(\alpha)}_a
g^{(\alpha)}_b}
{M^2_\alpha-s}+\Phi_{ab}(s).
\label{2}
\ee
The analysis of Ref. \cite{aps} evidently shows that five $K$-matrix 
poles are needed for fitting to data at $\sqrt {s} \leq 1900$ MeV. 
The $K$-matrix pole corresponds to a meson state with the switched off 
decay channels:  $g^{(\alpha)}_a \to 0$. However, the experimentally 
observed couplings $g^{(\alpha)}_a $ are not small, and these 
couplings, being responsible for the decay of meson states, are also 
responsible for a strong meson mixing.  Moreover, the masses of mixed 
states differ significantly from the primary ones. These "primary 
mesons" were called  "bare mesons" \cite{aps}, in contrast to physical 
states, for which the cloud of real particles ($\pi\pi$, $K\bar K$, 
$\eta\eta$, and so on) plays an important role in their formation.  
The coupling constants of bare state, $g^{\alpha}_a$, depend on 
quark/gluon content of the state $a$, so the restoration of  couplings 
for the channels $\pi\pi$, $K\bar K$, $\eta\eta$,  $\eta\eta'$   makes 
it possible to perform the $q\bar q$/gluonium classification of bare 
$00^{++}$ states.  The flavour content of the scalar/isoscalar bare 
states, members of the nonets $1^3P_0 q\bar q$ and $2^3P_0 q\bar q$, 
is determined by the mixing angle $\phi$ as follows: $q\bar q=n\bar 
n\cos\phi+s\bar s\sin\phi $ where $n\bar n =(u\bar u +d\bar d)/\sqrt 2 
$.

Concerning the non-pole term, a smooth $s$-dependence was allowed
for $\Phi_{ab}(s)$.

Two solutions for the bare
states in the mass region $\sqrt s\leq 1900$ MeV were found in Ref.
\cite{aps}.

{\bf Solution I}:
\be
\begin{array}{lll}
Classification & \;\;\; \;\;State & Mixing \; angle   \\
\; 1^3P_0 q\bar q & f_0^{\rm bare}(720\pm 100)  \;\;\;
& -69^\circ\pm 12^\circ \\
1^3P_0 q\bar q & f_0^{\rm bare}(1260\pm 30) \;\;\;& 21^\circ\pm
12^\circ \\ 2^3P_0 q\bar q & f_0^{\rm bare}(1600\pm 50) \;\;\; &
-6^\circ\pm 15^\circ \\ 2^3P_0 q\bar q & f_0^{\rm bare}(1810\pm 30)
\;\;\; & 84^\circ\pm 15^\circ \\
Gluonium & f_0^{\rm bare}(1235\pm 50) \;\;\; & ~ \\
\end{array}
\label{3}
\ee
{\bf Solution II}:
\be
\begin{array}{lll}
Classification & \;\;\; \;\;State & Mixing \; angle   \\
\; 1^3P_0 q\bar q & f_0^{\rm bare}(720\pm 100)  \;\;\;
& -69^\circ\pm 12^\circ \\
1^3P_0 q\bar q & f_0^{\rm bare}(1260\pm 30) \;\;\;& 21^\circ\pm
12^\circ \\ 2^3P_0 q\bar q & f_0^{\rm bare}(1235\pm 50) \;\;\; &
40^\circ\pm 10^\circ \\ 2^3P_0 q\bar q & f_0^{\rm bare}(1810\pm 30)
\;\;\; & -50^\circ\pm 10^\circ \\
Gluonium & f_0^{\rm bare}(1560\pm 30) \;\;\; & ~ \\
\end{array}
\label{4}
\ee
Both $K$-matrix solutions, I and II, lead to nearly  identical
positions of the amplitude poles in the complex mass plane.
The amplitude has five poles:
\be
\begin{array}{cl}
Resonance & Pole\; position \; (in \;MeV) \\
             ~& ~ \\
f_0(980)  & 1015\pm15-i(43\pm8)\\
f_0(1300) & 1300\pm20-i(120\pm20)\\
f_0(1500) & 1499\pm8-i(65\pm10) \\
f_0(1750) & 1750\pm30-i(125\pm70)\\
f_0(1530^{+90}_{-250}) & 1530^{+90}_{-250}-i(560\pm140)\ .
\end{array}
\label{5}
\ee
An appearence of the broad resonance  $f_0(1530^{+90}_{-250})$ is not
accidental: a large width of  $f_0(1530^{+90}_{-250})$ is due
to an effect of the accumulation of widths of neighbouring
resonances \cite{lock}. In both solutions, the broad state is
the descendant of the scalar gluonium keeping $\sim 50\%$ of its
component.

\subsection {$K$-matrix elements for the transitions
 $\gamma\gamma\to\pi\pi$ and $\gamma\gamma\to K \bar K$}

The $K$-matrix element for a prompt production of mesons,
$\widetilde K_{\gamma\gamma , a}$ of Eq. (\ref{1}), is
written in a form similar to Eq. (\ref{2}):
\be
\widetilde K_{\gamma\gamma , a}(s)=
 \sum_{\alpha =1}^4
\frac{F^{\bf{bare}}_{\gamma\gamma \to f_0(\alpha)}
g^{(\alpha)}_a}
{M^2_\alpha-s}+f_{\gamma\gamma ,a}.
\label{6}
\ee
Here $ F^{\bf{bare}}_{\gamma\gamma \to f_0(\alpha)}$ is the
form factor for the production of  bare scalar/isoscalar
$q\bar q$ state: $\gamma\gamma \to f^{\bf{bare}}_0 (\alpha)$.
The summation is carried  over all the $q\bar q$ states (Eq. (\ref{3})
for Solution I and Eq. (\ref{4}) for  Solution II). The term $
f_{\gamma\gamma ,a}$ refers to the background
contribution.

\subsection{Coupling $F^{\bf bare}_{f_0 \to \gamma\gamma}$}

In calculation of $F^{\bf bare}_{f_0\to \gamma\gamma}$,
we follow the
method developed in \cite{aabn,amn} where the form factors for
the transition
processes $meson \to \gamma^*(Q^2)\gamma$ were studied; in the
limit $Q^2 \to 0$, the transition form factor provides the decay
coupling $meson \to \gamma\gamma$ which is written in (\ref{6}).
The method is based on the double spectral
representation for the transition amplitudes and application  of the
light-cone wave functions for mesons involved. An important
point is that in the analysis of the reactions
 $\pi^0 \to \gamma^*(Q^2)\gamma$, $\eta \to
\gamma^*(Q^2)\gamma$ and $\eta' \to \gamma^*(Q^2)\gamma$
the vertex for the transition $\gamma \to q\bar q$ (or photon
wave function)
was determined in Ref. \cite{amn}.

Following the prescription of Ref. \cite{aabn}, we present
the couling $F^{\bf bare}_{f_0\to \gamma\gamma}$ in
terms of the light cone variables:
\beq
F^{\bf bare}_{f_0 \to \gamma\gamma}&=&
\frac {2 \sqrt {N_c} }{16\pi^3}
\int \limits_{0}^{1}
\frac {dx}{x(1-x)^2}  \int d^2k_{\perp}
\nonumber \\
&\times&
\bigg [\cos \phi \; Z_{n\bar n}T_{n\bar n}(x,\vec k_{\perp})
\Psi^{\bf bare}_{f_0} (M^2_{n\bar n})
\Psi_{\gamma \to n\bar n} (M^2_{n\bar n})
\nonumber \\
&+&\sin \phi \;
Z_{s\bar s}T_{s\bar s}(x,\vec k_{\perp})
\Psi^{\bf bare}_{f_0} (M^2_{s\bar s})
\Psi_{\gamma \to s\bar s} (M^2_{s\bar s})
\bigg ].
\label{7}
\eeq
Here
$x$ and $k^2_{\perp}$ are the light cone variables of quarks;
$\Psi^{\bf bare}_{f_0} (M^2_{q\bar q})$ is the $q\bar q$ wave function
of  $f^{\bf bare}_0$, and $M^2_{q\bar q}$ is
the $q\bar q$ invariant mass squared for strange, $M^2_{s\bar s}$,
or non-strange, $M^2_{n\bar n}$, quarks:
\be
M^2_{q\bar q}
=\frac{m^2+k^2_{\perp} }{x(1-x)}\,.
\label{8}
\ee
Here $m$ is the constituent quark mass;
$Z_{n\bar n} $ and  $Z_{s\bar s}$ are charge factors
for $n\bar n$ and $s\bar s$ components:
$Z_{n\bar n}=(e_u^2+e_d^2)/\sqrt 2  = 5 /9\sqrt 2$
and $Z_{s\bar s}=e_s^2 =1/9$.
The factor $\sqrt {N_c}$, where
$N_c=3$ is the number of colours, is related to the normalization of
the photon vertex; the photon wave functions
$\Psi_{\gamma \to n\bar n} (M^2_{n\bar n})$ and
$\Psi_{\gamma \to s\bar s} (M^2_{s\bar s})$ were
found in Ref. \cite{amn}.

The wave function  $f_0^{\bf bare}$ of the basic nonet $1^3P_0 q\bar q$
is parametrized in the exponential form:
\be
\Psi_{f_0}^{\bf bare}(M^2_{q\bar q})=
   Ce^{-bM^2_{q\bar q}},
\label{9}
\ee
where $C$ is normalization constant, and the parameter $b$ determines
the radius squared of the bare state.

The wave function of the first radial excitation,  $2^3P_0 q\bar q$,
is written in the exponential approximation as follows:
\be
\Psi_{f_0}^{{\bf bare\;(1)}}
(M^2_{q\bar q})=C_1(D_1M^2_{q\bar q} -1)e^{-b_1M^2_{q\bar q}}\,.
\label{10}
\ee
The parameter $b_1$ can be related to the radius of the radial
excitation state, then the values $C_1$ and $D_1$ are
fixed by the normalization and orthogonality requirements,
$\Psi_{f_0}^{{\bf bare\;(1)}}\otimes
\Psi_{f_0}^{{\bf bare\;(1)}}=1$ and
$\Psi_{f_0}      \otimes\Psi_{f_0}^{{\bf bare\;(1)}}=0$.

For the transition form factor,
the spin structure factor $T_{n\bar n}(x,\vec k_{\perp})$ is fixed by
the quark loop trace. It is equal to  \cite{aabn}:
\be
T_{q\bar q}(x,\vec k_{\perp}) =2mM^2_{q\bar q} \left [1+
\frac{k^2_{\perp}\cos^2 \varphi}
{M^2_{q\bar q} (1-x)^2 } \right ]^{-1} -8m^3,
\label{11}
\ee
where $\vec
k_{\perp} = (k_{\perp}\sin \varphi , \; k_{\perp} \cos \varphi )$.

\subsection{Background terms $f_{\gamma\gamma ,a}$ }

The background terms cannot be calculated unambiguously,
they are to be considered as free parameters.
One may suggest that dominant contribution into $f_{\gamma\gamma ,a}$
is given by pole diagrams with $t$- and $u$-channel singularities
located near the physical region:
these are diagrams of Fig. 1b type
with charged pion and kaon exchanges.
However, the
pole terms contain unknown $t$- and $u$-channel form factors:
in the pole diagram of Fig. 1b, the form factor refers to the
$\gamma\pi\pi$-vertex. These form factors cause
uncertainties in calculation of the pole diagram contributions
into $f_{\gamma\gamma ,a}$.

Below, when estimating $\pi\pi$ and $K\bar K$ spectra,
we follow an idea that  dominant contribution comes
from the processes with charge exchanges in the crossing $t$- and
$u$-channels. Therefore $f_{\gamma\gamma ,\pi^+\pi^-} \ne 0$ and
$f_{\gamma\gamma ,K^+K^-}\ne 0$, while all other $f_{\gamma\gamma
,a}$'s are equal to zero.
We parametrize
$f_{\gamma\gamma ,\pi^+\pi^-} $ in the form:
\be f_{\gamma\gamma
,\pi^+\pi^-} = \zeta_{\pi\pi} \frac{\pi}{\sqrt{3}}\int\limits_{-1}^{1}
dy\left [ 1+ \frac{s(s-4\mu^2)(1-y^2)}{s^2- s(s-4\mu^2)y^2} \right ]\,.
\label{12}
\ee
Here $\zeta_{\pi\pi} $ is a parameter, the factor $1/\sqrt 3$ is
 Clebsch-Gordon coefficient,
and $\mu$ is the pion mass.
The integral in (\ref{12}) is the $S$-wave projection of
the sum of the pion
exchange diagrams
in $t$- and $u$-channels: we
fix the $s$-dependent shape of $f_{\gamma\gamma
,\pi^+\pi^-} $ in a form which is given by
the closest pole terms in the $t$- and $u$-channels. The parameter
$\zeta_{\pi\pi} $ effectively accounts for the uncertainties
related to the contributions of form factors and  other more
distant diagrams.

The background term
$f_{\gamma\gamma ,K^+K^-} $ is parametrized by the
analogous expression,
with the replacements
$\zeta_{\pi\pi}/\sqrt 3 \to \zeta_{K\bar K}/\sqrt 2$ and $\mu \to
\mu_K$.

\section{Results}

The results of calculation of the $00^{++}$-wave two-meson spectra
in the reactions $\gamma\gamma\to\pi^0\pi^0$,
$\gamma\gamma\to\pi^+\pi^-$,
$\gamma\gamma\to K^0\bar K^0$ and
$\gamma\gamma\to K^+ K^-$ are shown in Figs. 2 to 5.

For the calculation of spectra, we fix the parameters $b$ and $b_1$ 
which determine the radii of  mesons $1^3P_0q\bar q$ and $2^3P_0q\bar 
q$, see Eqs. (\ref{9}) and (\ref{10}). The value $b$  is chosen to be 
the same as that of $a_0(980)$: following the results of Refs. 
\cite{aps,hep}, we suppose that $a_0(980)$ is a member of the $1^3P_0 
q\bar q$ nonet.  The calculation of  partial width $a_0(980) \to 
\gamma\gamma$ gives \cite{aabn}: $R^2(1^3P_0q\bar q)\sim 15$ 
GeV$^{-2}$.  The quark model teaches us that the meson radius for the 
first radial excitation should be greater than that of the basic 
nonet; correspondingly, we put $R^2(2^3P_0q\bar q)=22$ GeV$^{-2}$. In 
terms of a pion radius, it means that $R(1^3P_0q\bar q)/R_{\pi} \simeq 
1.22$ and $R(1^3P_0q\bar q)/R_{\pi} \simeq 1.48$.  The couplings 
$F^{\bf bare}_{f_0(\alpha) \to \gamma\gamma} $ are calculated for 
these values of radii.

The parameters $\zeta_{\pi\pi}$ and $\zeta_{K\bar K}$
define the background in the
$\pi\pi$ and $K\bar K$ spectra.
We use the data from the low-energy region of spectra
$\gamma\gamma \to \pi\pi$
(where the $S$-wave dominates) to fix $\zeta_{\pi\pi}$.
However, in this region the data of different groups differ
significantly from each other, so we present in Figs. 2 to 5 the
calculation results with $\zeta_{\pi\pi} = 0.5$  and
$\zeta_{\pi\pi} =0.35$ which provide better description of either
the data of Ref. \cite{a0f0} or Ref. \cite{a0}, correspondingly.

A relative sign of the wave functions
$\Psi_{f_0}^{{\bf bare}}$ and
$\Psi_{f_0}^{{\bf bare\;(1)}} $
is not fixed.  Two curves in Figs.  2 to 5, solid and dotted ones,
correspond to opposite signs of $C_1$.

In all calculation variants the resonances $f_0(980)$ and $f_0(1500)$ 
reveal themselves as dips in the $\pi^0\pi^0$ and $\pi^+\pi^-$ 
spectra; the resonance $f_0(1300)$ is hardly seen producing rather 
weak variations in the spectra. At $E > 1000$ MeV, the contribution of 
the $02^{++}$-wave increases being dominant in the region of the 
$f_2(1270)$-resonance.

For the $K\bar K$ spectra, we consider the variants with
$\zeta_{\pi\pi}=0.5$, $\zeta_{K\bar K}=0.7$ (Figs.  4a,b and 5a,b) and
$\zeta_{\pi\pi}=0.35$, $\zeta_{K\bar K}=0.44$ (Figs. 4c,d and 5c,d).
The variant
$\zeta_{\pi\pi}=0.5$, $\zeta_{K\bar K}=0.7$
supposes that the contribution
of the $02^{++}$-wave into $K^+K^-$ spectrum is small at
$1200$-$1400$ MeV. More realistic is the variant with
$\zeta_{\pi\pi}=0.35$, $\zeta_{K\bar K}=0.44$ where the peak in the
region $1300$ MeV is due to the $f_2(1270)$-resonance.  The
$00^{++}$-wave gives a small contribution to the $K^0\bar K^0$
spectrum, see Fig. 5, (the resonance
$a_0(980)$ can dominate in the region near $1000$ MeV).

\section{Conclusion}

The extraction of partial widths $f_0(980) \to \gamma\gamma$ and 
$f_0(1500) \to \gamma\gamma$ from studying the reactions $\gamma\gamma 
\to \pi\pi$ and $\gamma\gamma \to K\bar K$ requires detailed partial 
wave analysis of meson spectra. Here the situation is similar to that 
in the spectra $\pi \pi \to \pi\pi$: the resonances  $f_0(980)$ and 
$f_0(1500)$ reveal themselves as dips in both reactions.

The $\pi\pi$ and $K\bar K$ spectra are not sensitive to the position 
of a pure gluonium: solutions I and II, which refer to pure gluonium 
at 1230 and 1600 MeV, correspondingly, give rather similar spectra. 
The unification of the spectra with different positions of the 
gluonium is due to the final state interaction effects.

We thank D.V. Bugg, L.G. Dakhno, V.A. Nikonov and A.V. Sarantsev
for useful discussions and comments.
The paper was supported by grants RFFI 98-02-17236 and
INTAS-RFBR 95-0267.

\begin{figure}[h]
%fig.1
\centerline{\epsfig{file=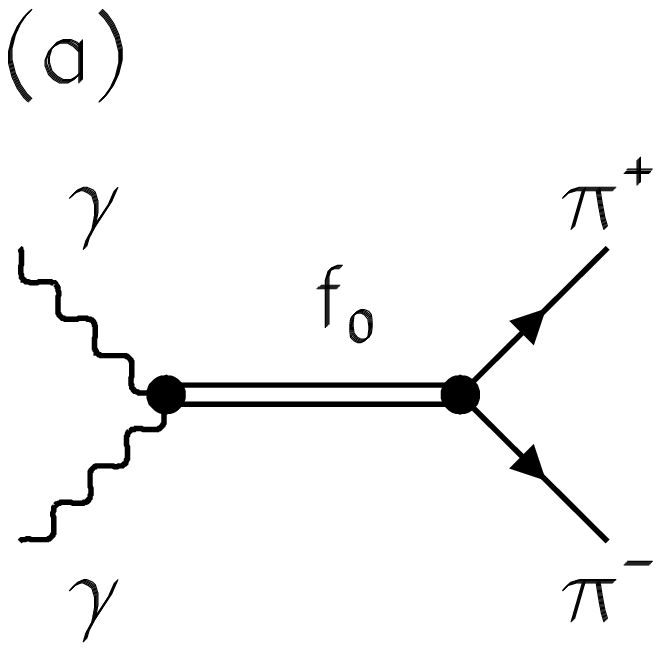,width=4.0cm}\hspace{1cm}
            \epsfig{file=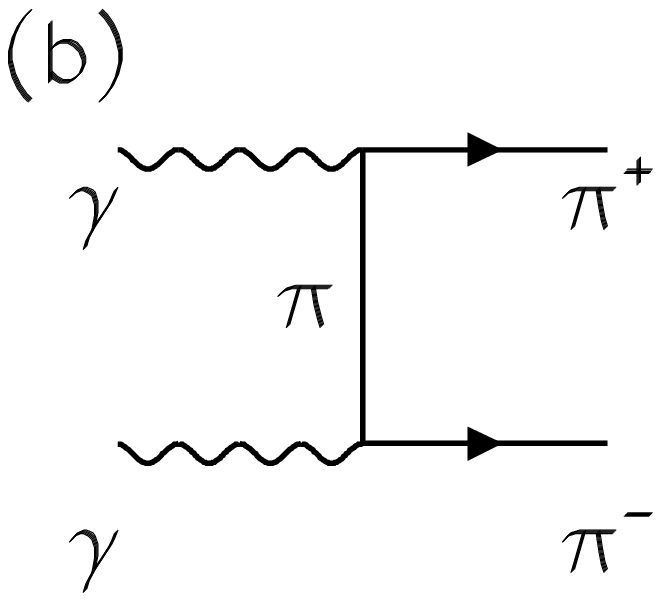,width=4.0cm}}
\caption{
Diagrams for the prompt production of $\pi^+ \pi^-$:
a) $s$-channel $f_0$-meson production,
b) $t$-channel (or $u$-channel) $\pi$-meson exchange process.
} \end{figure}

\begin{figure}
%Fig. 2
\centerline{\epsfig{file=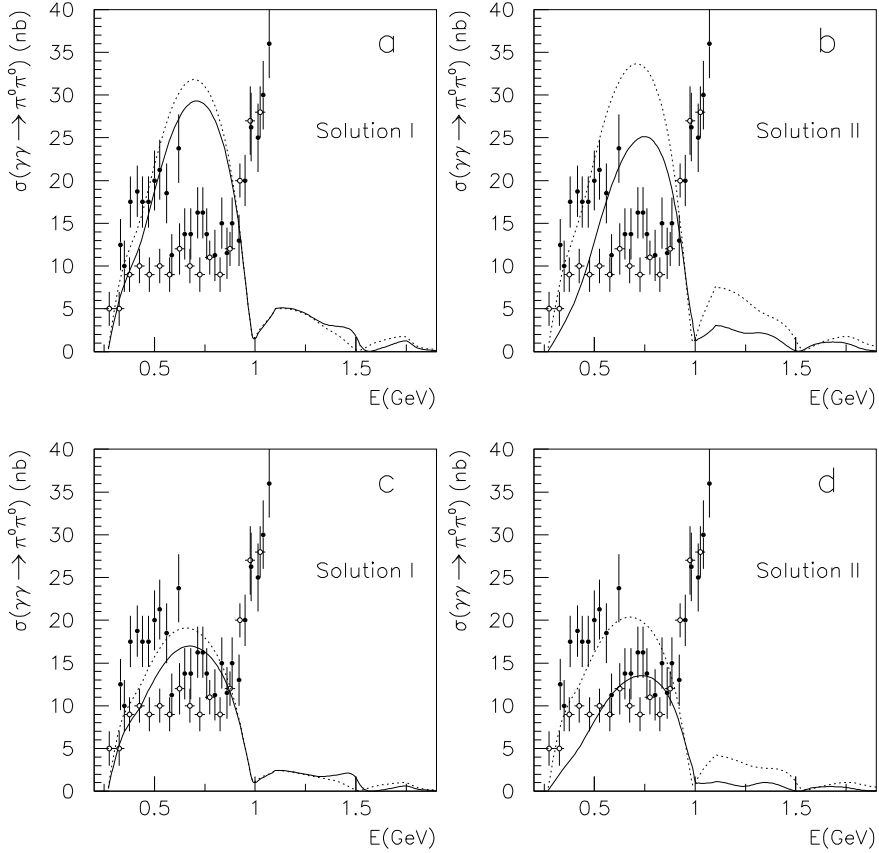,width=13cm}}
\caption {
The $00^{++}$ cross section for $\gamma\gamma \rightarrow
\pi^0\pi^0$ reaction. Solid and dashed curves correspond
to positive and negative signs of the normalization constant
for the first radial excitation wave function;
$\zeta_{\pi\pi}=0.5$ stands for (a) and (b), and
$\zeta_{\pi\pi}=0.35$  for (c) and (d).
Experimental data are taken from Refs. [2,4]: the data
include contributions of all waves.}
\end{figure}

\begin{figure}
%Fig. 3
\centerline{\epsfig{file=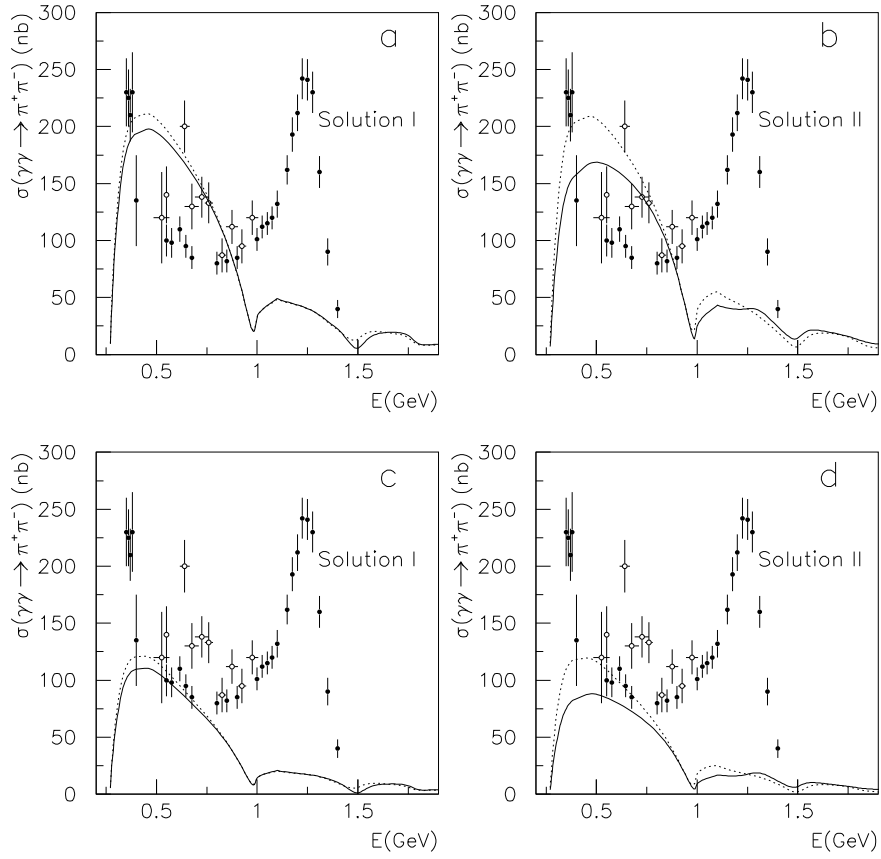,width=13cm}}
\caption {
The $00^{++}$ cross section for  $\gamma\gamma \rightarrow
\pi^+\pi^-$ reaction.
Notations are the same as for Fig. 2.
}
\end{figure}

\begin{figure}
%Fig. 4
\centerline{\epsfig{file=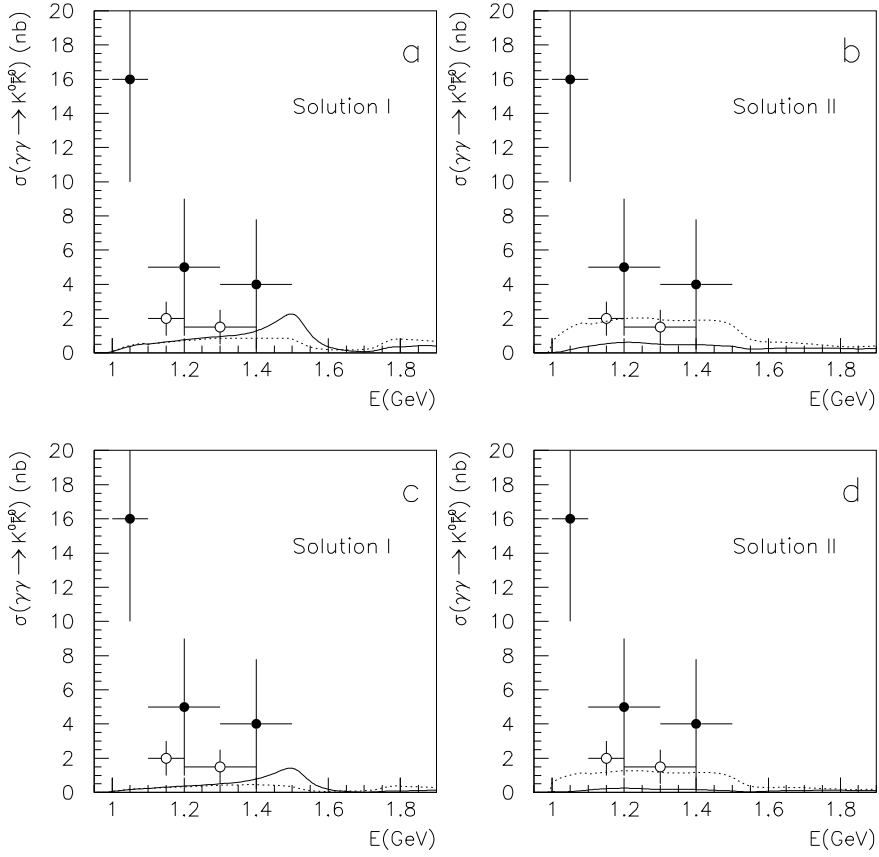,width=13cm}}
\caption {
The $00^{++}$ cross section for  $\gamma\gamma \rightarrow
 K^0 \bar K^0$ reaction. Curves are marked as in Fig.
2;  (a) and (b) correspond to $\zeta_{\pi\pi}=0.5$, $\zeta_{K\bar
K}=0.7$ and (c) and (d) correspond to $\zeta_{\pi\pi}=0.35$,
$\zeta_{K\bar K}=0.44$.
}
\end{figure}

\begin{figure}
%Fig. 5
\centerline{\epsfig{file=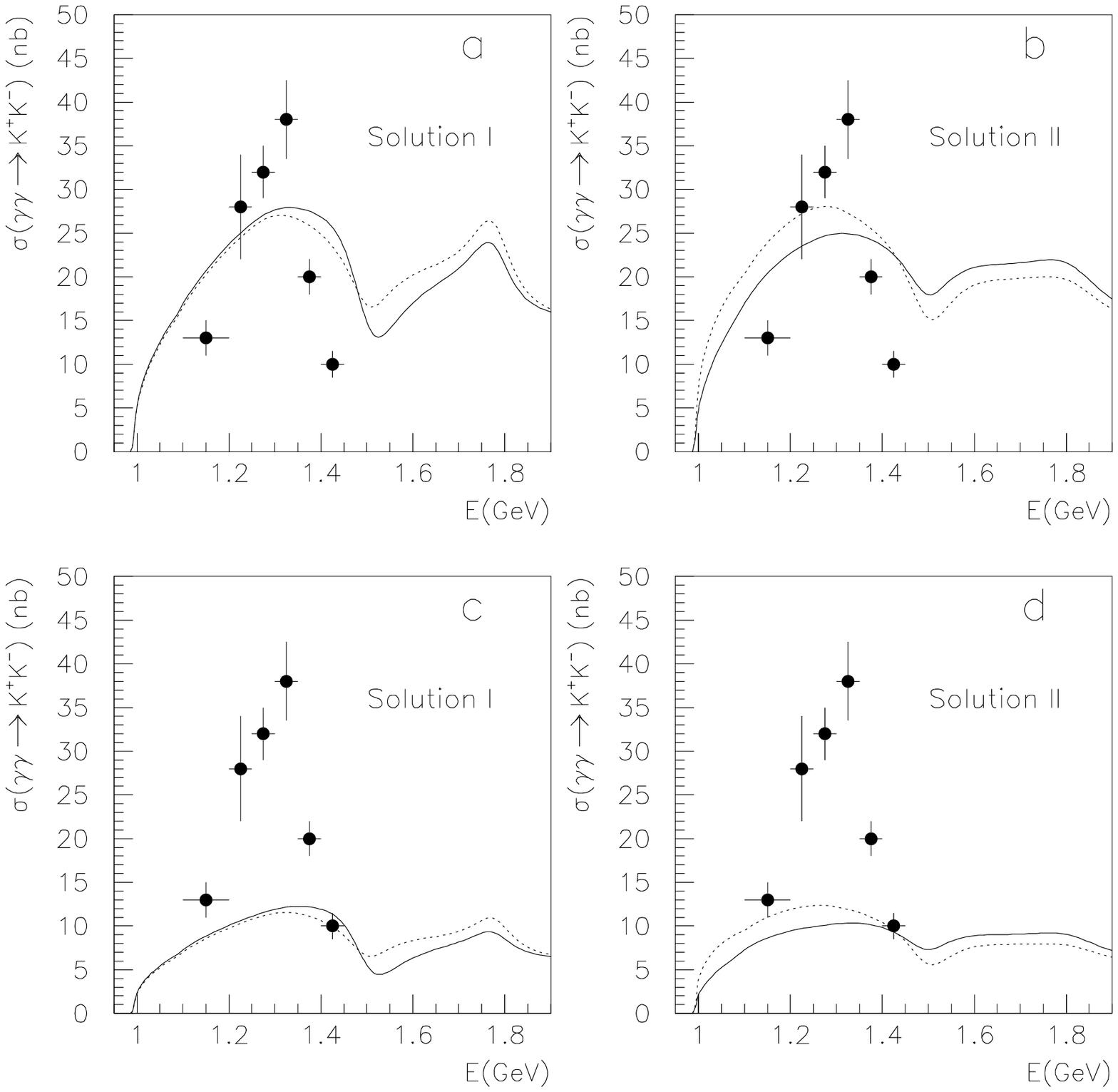,width=13cm}}
\caption {The $00^{++}$ cross section for  $\gamma\gamma \rightarrow
K^+K^-$ reaction. Notations are the same as in Fig. 4.}
\end{figure}


\begin{thebibliography}{99}
\bibitem{barnes} T. Barnes, {\it Theoretical Aspects of Light
Meson Spectroscopy}, in: Hadron Spectroscopy
and the Confinement Problem, Edited by D.V. Bugg,
Plenum Press, New York (1996);\\
 E. Klempt, {\it Hadron '97 Summary}, in:
Hadron Spectroscopy, AIP Conference Proceedings 432, Editors
S.-U. Chung and H.J. Willutzki, Woodbury, New York (1998).
\bibitem{a0f0} T. Oest et al. (JADE Coll.) Z. Phys. {\bf C47}
 (1990) 343.
\bibitem{f0} D. Morgan, M. Pennington, Z. Phys. {\bf C48} (1990)
623.
\bibitem{a0} D. Antreasyan et al. (Crystal Ball Coll.)
Phys. Rev. {\bf D33} (1986) 1847;\\
H. Mariske et al. (Crystal Ball Coll.)
Phys. Rev. {\bf D41} (1990) 3324.
\bibitem{PDG} C. Caso et al. (PD Group), Eur. Phys. J.
{\bf C3}, (1998) 1.
\bibitem{aabn}  A.V. Anisovich, V.V. Anisovich, D.V. Bugg, V.A.
Nikonov, {\it Partial widths
 $a_0(980) \to \gamma\gamma$, $f_0(980) \to \gamma\gamma$ and
$q\bar q$ classification of the lightest scalar mesons},
hep-ph 9903396,
Phys. Lett. {\bf B}, in press.
\bibitem{aps} V.V. Anisovich, Yu.D. Prokoshkin,
A.V. Sarantsev, Phys. Lett. {\bf B389} (1996) 388.
\bibitem{hep} V.V. Anisovich, A.A. Kondashov, Yu.D. Prokoshkin et al.
{\it Two-pion spectra for the reaction $\pi^- p\to \pi^0\pi^0 n $ at
38 GeV/c pion momentum and combined analysis of the GAMS,
Crystal Barrel and BNL data}, hep-ph/9711319, unpublished.
\bibitem{gams} D. Alde et al. (GAMS),
Zeit. Phys. {\bf C66} (1995) 375; \\
Yu. D. Prokoshkin et al. (GAMS), Physics-Doklady {\bf 342}
 (1995) 473; \\
 F. Binon et al. (GAMS), Nuovo Cim. {\bf A78}, 313 (1983);
  {\bf A80} (1984) 363.
\bibitem{bnl} S. J. Lindenbaum and R. S. Longacre, Phys. Lett.
{\bf B274} (1992) 492; \\
A. Etkin et al., Phys. Rev. D {\bf 25} (1982) 1786.
\bibitem{cbc} V.V. Anisovich et al. (Crystal Barrel Coll.),
 Phys. Lett. {\bf B323} (1994)  233;\\
 C. Amsler et al. (Crystal Barrel Coll.), Phys. Lett. {\bf B342}
 (1995) 433; {\bf B355} (1995) 425.
\bibitem{lock} V.V. Anisovich, D.V. Bugg, A.V. Sarantsev,
Phys. Rev. {\bf D 58:}111503 (1998).
\bibitem{amn} V.V. Anisovich, D.I. Melikhov, V.A. Nikonov,
Phys. Rev. {\bf D55} (1997) 2918 ; Phys. Rev. {\bf D52} (1995) 5295;\\
V.V. Anisovich, D.V. Bugg, D.I. Melikhov, V.A.
Nikonov, Phys. Lett. {\bf B404} (1997) 166.

\end{thebibliography}
\end{document}